\input{aipcheck}

\documentclass[
    ,final            
  ]
  {aipproc}

\layoutstyle{6x9}

\usepackage{amsmath}
\usepackage{bm}

\begin{document}

\title{Instabilities in two flavor quark matter}

\classification{12.38.-t,~74.20.Fg} \keywords{High density QCD,
Nambu-Jona Lasinio models, Instabilities in quark matter.}

\author{Marco Ruggieri}{
  address={Dipartimento di Fisica, Universit\`a di Bari, I-70126
Bari, Italy \\ I.N.F.N., Sezione di Bari, I-70126 Bari, Italy}}

\begin{abstract}
I discuss briefly the instabilities of two flavor quark matter,
paying attention to the gradient instability which develops in the
g2SC phase in the Goldstone $U(1)_A$ sector.
\end{abstract}

\maketitle


It is widely accepted that at high density and low temperature, the
ground state of deconfined quark matter is the color
superconductor~\cite{Alford:1997zt,Rapp:1997zu,Alford:1998mk}. In
this talk I consider two flavor superconductive deconfined quark
matter, which consists of $u$ and $d$ massless quarks, whose action
is given by
\begin{equation}
S = \int d^4x \left[\bar\psi_{i\alpha}\left(i
\gamma^{\mu}\partial_\mu \delta^{\alpha\beta}\delta_{ij}+
\mu_{ij}^{\alpha\beta}\gamma_0\right)\psi_{j\beta} + (L\rightarrow
R) + {\cal L}_\Delta\right]~.
\end{equation}
In the above equation, $\psi$ denotes left-handed fields; Greek
(Latin) indices stem for color (flavor); $\mu_{ij}^{\alpha\beta}$
is the chemical potential matrix, depending on the mean quark
chemical potential $\mu$, the charge chemical potential $\mu_Q =
-\mu_e$ and the color chemical potentials $\mu_3$, $\mu_8$.
Condensation in the quark-quark channel is described by the
lagrangian ${\cal L}_\Delta$ which is given in the mean field
approximation by
\begin{equation}
{\cal L}_\Delta = -\frac{\Delta}{2}\psi_{i\alpha}^T C
\psi_{j\beta}\epsilon^{\alpha\beta3}\epsilon_{ij} + H.c.
-(L\rightarrow R)~,\label{eq:LagrGap}
\end{equation}
with $C$ the charge conjugation matrix. We assume that in the
ground state
\begin{equation}
\langle\psi_{i\alpha}^L C\psi_{j\beta}^L\rangle =
-\langle\psi_{i\alpha}^R C\psi_{j\beta}^R\rangle
\propto\Delta\epsilon^{\alpha\beta3}\epsilon_{ij}\neq0~,\label{eq:VEV}
\end{equation}
where the superscripts $L,R$ denote left-handed and right-handed
quarks respectively. Eq.~\eqref{eq:VEV} means that pairs $u_r -
d_g$, $u_g - d_r$ are formed, with zero total momentum and zero
total spin; on the other hand the blue quarks do not have a role in
the pairing phenomenon.

The quark chemical potential matrix is given by
\begin{equation}
{\bm \mu} = \left(\mu{\bm 1}_F - {\bm Q}\mu_e\right)\otimes{\bm 1}_C
+ {\bm 1}_F \otimes\left(\mu_3 {\bm T_3} + \mu_8 {\bm T_8}\right)~.
\end{equation}
In the ground state described by Eq.~\eqref{eq:VEV} one has $\mu_3
= 0$ and $\mu_8 = {\cal O}(\Delta^2/\mu)$; on the other hand,
$\mu_e = {\cal O}(0.1\mu) \gg \mu_8$. Therefore in what follows we
assume $\mu_8 = 0$ in order to simplify the calculations. We
stress that even considering $\mu_8\neq0$ does not change the
results presented here, since the eight color chemical potential
does not change the difference of the chemical potentials between
the paired quarks, which are the relevant ones in this context.
With this choice one has $\mu_u = \bar\mu - \delta\mu$ and $\mu_d
= \bar\mu + \delta\mu$, with $\delta\mu = \mu_e/2$. Hence, once
$\bar{\mu}$ is fixed, only $\delta\mu$ is needed to specify the
spectrum of the system. The dispersion laws of the paired quarks
are
\begin{equation}
E_{\pm\pm} = \left|\pm\delta\mu\pm\sqrt{(p-\bar\mu)^2 +
\Delta^2}\right|~, \label{eq:P}
\end{equation}
where $\bar\mu$ is the mean chemical potential and $\Delta$ is the
gap parameter. It is easily realized that if $\delta\mu > \Delta$
then the dispersion law~\eqref{eq:P} has two nodes, and is thus
gapless. When $\delta\mu < \Delta$ the phase is the 2SC; on the
other hand if $\delta\mu > \Delta$ the phase is the gapless 2SC
(g2SC) phase, considered in the QCD context
in~\cite{Shovkovy:2003uu}.

Neglecting electromagnetism, the symmetry group of two massless
flavor QCD at high chemical potential (axial symmetry is unbroken at
high density),
\begin{displaymath}
G_{QCD} = SU(3)_c \otimes U(2)_V \otimes U(2)_A~,
\end{displaymath}
is broken down by the quark condensate $\langle\psi\psi\rangle$ to
\begin{displaymath}
G_{2SC} = SU(2)_c \otimes U(2)_V \otimes SU(2)_A~.
\end{displaymath}
There is a broken $U(1)_A$ since the diquark is not invariant for a
phase shift of the quark fields. Then, from the Goldstone theorem it
follows that one  massless scalar appears in the spectrum,
corresponding to the breaking of $U(1)_A$ . The color group is also
broken: as a consequence five of the eigth gluons become massive
(Meissner effect, familiar from ordinary superconductivity). In this
talk I focus on the Goldstone mode related to $U(1)_A$, discussing
briefly the Meissner effect.

The Goldstone field $\phi$ describes small fluctuations of the
condensate around its mean field value. It can be introduced in the
model by the replacement, in the quark lagrangian,
$\langle\psi\psi\rangle \rightarrow
e^{2i\phi/f}\langle\psi\psi\rangle$. Integration over the quark
fields in the functional integral gives rise to the  one loop
effective action of $\phi$; it can be written in the low energy
regime $p \ll \Delta$ as~\cite{Gatto:2007ja}
\begin{equation}
{\cal L}(p) = \frac{1}{2}\left[p_0^2 \phi^2 - v^2 ({\bm p}\phi)
\cdot ({\bm p}\phi)\right]~.
\end{equation}
Evaluation of the loop integrals gives
\begin{displaymath}
f^2 = \frac{4\mu^2}{\pi^2}\left(1 -
\theta(\delta\mu-\Delta)\frac{\sqrt{\delta\mu^2 -
\Delta^2}}{\delta\mu}\right)~,
\end{displaymath}
obtained by the requirement of canonical normalization of the
lagrangian, and
\begin{equation}
v^2 = \frac{1}{3}\theta(\Delta-\delta\mu) -
\frac{1}{3}\theta(\delta\mu-\Delta)\frac{\delta\mu}{\sqrt{\delta\mu^2
- \Delta^2}}~. \label{eq:vSq}
\end{equation}
Eq.~\eqref{eq:vSq} shows that $v^2 < 0$ in the g2SC phase. We thus
have a {\sl gradient instability} of $\phi$.

The condensate in Eq.~\eqref{eq:VEV} breaks $SU(3)_c$ down to
$SU(2)_c$: as a consequence, five of the eight gluons are massive
(Higgs mechanism). In particular, one can evaluate the Meissner
masses, defined by $m_M^2 = - \Pi(p_0 = 0,{\bm p}\rightarrow 0)$
where $\Pi$ is the polarization tensor of the gluons. One
finds~\cite{Huang:2004bg}
\begin{eqnarray}
m_{M,4}^2 &=&
\frac{4\alpha_s\mu^2}{3\pi}\left(\frac{\Delta^2-2\delta\mu^2}{2\Delta^2}
  + \frac{\delta\mu\sqrt{\delta\mu^2 - \Delta^2}}{\Delta^2}\theta(\delta\mu-\Delta)    \right)~,
\\
m_{M,8}^2 &=&
\frac{4\alpha_s\mu^2}{9\pi}\left(1-\frac{\delta\mu}{\sqrt{\delta\mu^2
- \Delta^2}}\right)~.
\end{eqnarray}
One notices the relation $f^2 v^2 \propto m_{M,8}^2$, linking the
instability in the Goldstone sector to the one in the gluon sector.
Moreover we notice that in the interval
$(\Delta/\sqrt{2})\leq\delta\mu\leq\Delta$ an instability in the
sector of the gluons $a=4,\dots,7$ occurs. However it cannot be
related to the Goldstone velocity instability since in this interval
$v^2 > 0$. This is an evidence of the different nature of the
instabilities between the gluons with $a=4,\dots,7$ and the gluon
with $a=8$.

Beside Meissner instability, the dispersion laws of dynamical gluons
have been studied~\cite{Gorbar:2006up}: it was found that the gluons
with $a=4,\dots,7$ have a negative squared plasmon mass for
$\delta\mu \leq \Delta/\sqrt{2}$ and a positive squared velocity. On
the other hand, the gluon with $a=8$ is massless and has a negative
squared velocity. It is interesting to notice that these kind of
instabilities are found both for the electric and for the magnetic
gluon modes, while the Meissner instability develops only for the
magnetic gluons.

The result $v^2 < 0$ in Eq.~\eqref{eq:vSq} can be interpreted as
$\langle{\bm\nabla}\phi\rangle \neq 0$. From the ansatz
\begin{equation}
\phi(t,{\bm x}) = {\bm\Phi}\cdot{\bm x} + h(t,{\bm
x})~,\label{eq:p1}
\end{equation}
and assuming $\langle{\bm\nabla}h\rangle \neq 0$~\footnote{The
assumption $\langle{\bm\nabla}h\rangle \neq 0$ is justified {\sl a
posteriori} since we find that the squared velocity of the
fluctuation field $h$ is positive, see Eq.~\eqref{eq:vivj}.} we get
$\langle{\bm\nabla}\phi\rangle = {\bm\Phi}$, and we call $\bm{\Phi}$
the Goldstone current. The bosonization of the quark lagrangian is
done via the transformation $\langle\psi\psi\rangle \rightarrow
e^{2i{\bm \Phi}\cdot{\bm x}/f}\langle\psi\psi\rangle e^{2ih/f}$. The
value $\Phi_0$ of $|{\bm\Phi}|$ in the ground state is evaluated by
minimizing the thermodynamic potential $\Omega$. Expanding the quark
propagator in powers of $\Delta/q$ with $q \equiv |{\bm\Phi}|/f$ we
get $\Phi_0^2 \approx 1.2 \times f^2 \delta\mu^2$. Moreover,
expanding in the small field $h/f$ and evaluating the quark loops we
find
\begin{equation}
{\cal L}[h] = \frac{1}{2}\left((\partial_0 h)^2 - v_i v_j \partial_i
h \partial_j h  \right)~,\label{eq:d}
\end{equation}
and the low energy parameters are $(\Phi_x = \Phi_y = 0, \Phi_z =
\Phi_0)$
\begin{equation}
f^2 \approx 0.46 \mu^2\Delta^2 /\delta\mu^2~,~~~~~ {\bm v} =
(0,0,1)~.\label{eq:vivj}
\end{equation}
Therefore the tensor $v_i v_j$ is semidefinite positive, signaling
the Goldstone stability of the new ground state. The breaking of the
rotational symmetry due to ${\bm\Phi} \neq 0$, $SO(3)\rightarrow
SO(2)$, is reflected by the anisotropy of the tensor $v_i v_j$.

The phase shift in the quark lagrangian, in the case of ${\bm\Phi}
\neq 0$, is equivalent to consider an inhomogeneous
superconductive state with gap $\Delta e^{2i{\bm \Phi} \cdot {\bm
x}/f}$: this is the one-plane-wave (1PW) LOFF
state~\cite{Alford:2000ze}, where the fluctuation field $h$ plays
the role of the $U(1)_A$ Goldstone mode. From the previous results
we know that the Goldstone mode in the one-plane-wave state does
not suffer the velocity instability. Moreover, the Meissner masses
of the gluons in the 1PW state are positive (at least for small
$\Delta/\delta\mu$, see below). But now we are faced with a
problem: the stability criterion is a necessary but not a
sufficient condition for the existence of a given phase, since one
needs to compute the free energy in order to establish if this
state is the ground state or not. Since there is an equivalence
between the state with the Goldstone current (with ansatz given
by~\eqref{eq:p1}) and the one-plane-wave LOFF state, the free
energy of the two phases is the same. From the LOFF literature we
know that the free energy of the 1PW state yields it to be the
ground state of two flavor quark matter, in the weak coupling
limit, if
\begin{displaymath}
0.707\Delta_0\leq\delta\mu\leq 0.754\Delta_0~,~~~\Delta_0 \equiv
\Delta(\delta\mu = 0)~.
\end{displaymath}
This is a very narrow range. As a consequence, although the 1PW
state satisfies the stability conditions, its free energy does not
allow for the existence of this state in a wide interval of
$\delta\mu$. Hence the 1PW state is unlikely to be the ground
state of QCD. At this point, since we need a new state that is
stable and has a lower free energy, we can either look for
different a ansatz of the Goldstone current, or improve the 1PW
state by adding more plane waves. The latter case is easier to be
treated, although in this case we lose the correspondence with the
Goldstone current state. We follow the latter philosophy, leaving
the search for a different current ansatz to a future project.

In the multiple-plane-wave (MPW) state the ansatz for the gap is
given by~\cite{Bowers:2002xr,Casalbuoni:2004wm}
\begin{equation}
\langle\psi\psi\rangle \propto \Delta \sum_{a=1}^P e^{2i{\bm q}^a
\cdot {\bm x}}~.\label{eq:o}
\end{equation}
The low energy effective action for the $U(1)_A$ Goldstone mode is
obtained following the same procedure adopted in the case of the
1PW state. The result in configuration space is again given by
Eq.~\eqref{eq:d}, with the squared velocity tensor being defined
at the leading order in $\Delta/q$ as~\cite{Gatto:2007ja} (see
also~\cite{Mannarelli:2007bs} for a similar calculation)
\begin{equation}
v_i v_j = \sum_{a=1}^P (\hat{q}^a)_i (\hat{q}^a)_j/P~;
\end{equation}
it is an easy task to prove that it is enough to consider a
structure with three orthogonal wave vectors in order to yield a
definite positive tensor $v_i v_j$. Hence the Goldstone stability
requirement is fulfilled in the MPW state. For example we find for
the 1PW state
\begin{equation}
v_i v_j = [\text{diag}(0,0,1)]_{ij}~,~~~~~\text{1PW},
\end{equation}
while for the FCC structure (corresponding to $P=8$ in
Eq.~\eqref{eq:o}, with the wave vectors pointing to the edges of a
cube) we find
\begin{equation}
v_i v_j = \frac{1}{3}\delta_{ij}~,~~~~~\text{MPW}.
\end{equation}
Moreover, the Meissner stability requirement is also satisfied in
the MPW state. As a matter of fact, at the leading order in
$\Delta/q$ we find~\cite{Gatto:2007ja,Giannakis:2005vw},
\begin{eqnarray}
\left({\cal{M}}_{44}^{ij}\right)^2 &=& \frac{f^2}{16}v_i
v_j~,\\
\left({\cal{M}}_{88}^{ij}\right)^2 &=& \frac{f^2}{12}v_i v_j~.
\end{eqnarray}
and ${\cal{M}}_{55}^{ij} = {\cal{M}}_{66}^{ij} = {\cal{M}}_{77}^{ij}
= {\cal{M}}_{44}^{ij}$. It is evident that the positivity of the
Meissner tensor $\left({\cal{M}}_{ab}^{ij}\right)^2$ follows from
the positivity of the tensor $v_i v_j$.

We discuss now the interval of $\delta\mu$ in which the LOFF state
is stable respect to the normal phase. We have found that the BCC
structure ($P=6$, the wave vectors pointing to the faces of a
cube) is stable in the interval
$0.71\Delta_0\leq\delta\mu\leq0.95\Delta_0$; on the other hand the
FCC structure is favored in the range
$0.95\Delta_0\leq\delta\mu\leq1.35\Delta_0$. Therefore LOFF state
in the MPW configuration is stable in the interval
\begin{equation}
0.71\leq\frac{\delta\mu}{\Delta_0}\leq1.35~.\label{eq:interval}
\end{equation}
For larger values of $\delta\mu/\Delta_0$ the LOFF state is no
longer energetically favored, and cannot solve the instability
problem of high density QCD.

In conclusion, I have shown that beside chromomagnetic
instabilities, homogeneous gapless color superconductive quark
matter suffers of a Goldstone instability too, related to the
negative squared velocity of the Goldstone mode. My results suggest
that inhomogeneous superconductive states can solve this problem, if
$\delta\mu$ lies in the interval~\eqref{eq:interval}. Beside them,
gluonic phases~\cite{Gorbar:2005rx,Gorbar:2007vx,Hashimoto:2007ut}
may have a relevant role to solve the instability puzzle of high
density QCD. It should be noticed that nowadays  a comparison of the
free energies of the LOFF state (in the MPW state) and of the
gluonic phases is still lacking (for studies about the comparison of
the free energies of the one plane wave and the gluon phases
see~\cite{Kiriyama:2006ui}). My results suggest that such a
comparison should be done as soon as possible, as from it we could
learn much more we know about the ground state of neutral
superconductive quark matter.


\begin{theacknowledgments}
I wish to thank R.~Gatto and G.~Nardulli  for the fruitful
collaboration.
\end{theacknowledgments}

\end{document}